%% file: main_acm.tex
\documentclass[sigconf,screen,nonacm]{acmart}

\setcopyright{none}
\AtBeginDocument{%
  }

\usepackage{amsmath,amsfonts}
\usepackage{graphicx}
\usepackage{textcomp}
\usepackage{url}
\usepackage{subcaption} 
\usepackage{booktabs}
\usepackage{multirow}
\usepackage{makecell}
\usepackage{xspace}
\usepackage{balance} 
\usepackage[ruled,vlined,linesnumbered]{algorithm2e}
\usepackage{enumitem}

\newcommand\dbrepro{\ensuremath{\emph{DBRepro}}\xspace}
\newcommand\rsgen{\ensuremath{\textit{RSGen}}\xspace}
\newcommand\mirage{\ensuremath{\textit{Mirage}}\xspace}
\newcommand\touchstone{\ensuremath{\textit{Touchstone}}\xspace}
\newcommand\hydra{\ensuremath{\textit{Hydra}}\xspace}

\begin{document}

\title{DBRepro: Automated Database Synthesis via a Hybrid Constraint-Solving Approach for Reproducing Slow Queries}


\author{Zhaoyang Zhang}
\orcid{0009-0005-6892-7739}
\affiliation{%
  \institution{Renmin University of China}
  \city{Beijing}
  \country{China}
}
\email{zhaoyangzhang@ruc.edu.cn}

\author{Shuang Liu}
\correspondingauthor
\orcid{0000-0001-8766-7235}
\affiliation{%
  \institution{Renmin University of China}
  \city{Beijing}
  \country{China}
}
\email{Shuang.Liu@ruc.edu.cn}

\author{Dengfeng Xu}
\orcid{0009-0005-0675-4967}
\affiliation{%
  \institution{China Electronics Technology Kingbase (Beijing) Technologies Inc.}
  \city{Beijing}
  \country{China}
}
\email{xudengfeng@kingbase.com.cn}

\author{Wei Lu}
\orcid{0000-0001-6769-2695}
\affiliation{%
  \institution{Renmin University of China}
  \city{Beijing}
  \country{China}
}
\email{lu-wei@ruc.edu.cn}

\author{Jianquan Leng}
\orcid{0009-0006-3885-7829}
\affiliation{%
  \institution{China Electronics Technology Kingbase (Beijing) Technologies Inc.}
  \city{Beijing}
  \country{China}
}
\email{jqleng@kingbase.com.cn}

\author{Sheng Du}
\orcid{0009-0007-8355-8630}
\affiliation{%
  \institution{China Electronics Technology Kingbase (Beijing) Technologies Inc.}
  \city{Beijing}
  \country{China}
}
\email{dusheng@kingbase.com.cn}

\author{Xiaoyong Du}
\orcid{0000-0002-5757-9135}
\affiliation{%
  \institution{Renmin University of China}
  \city{Beijing}
  \country{China}
}
\email{duyong@ruc.edu.cn}

\begin{abstract}

Slow queries frequently cause severe performance bottlenecks in Database Management Systems (DBMSs). Due to the severe risk of exacerbating online resource contention, diagnosing their root causes online is usually infeasible; offline diagnosis has thus become the de facto approach. However, strict data privacy regulations prohibit accessing original user data. Therefore, synthesizing a proxy database, one that induces the query optimizer to generate identical physical execution plans as in production, from non-intrusive metadata, is critical for reproducing and diagnosing slow queries offline. Achieving high-fidelity reproduction demands satisfying both global statistical distributions and exact local cardinality constraints. Existing works fundamentally fail to achieve both simultaneously: data-driven approaches cannot enforce strict local cardinalities, whereas workload-aware approaches severely distort the overall data distribution, misleading the optimizer into divergent execution plans. To bridge this gap, we introduce \dbrepro, an automated end-to-end database synthesis framework that synergistically integrates data-driven and workload-aware methodologies. \dbrepro\ formulates database generation as a constrained distribution synthesis problem. It initializes a global data distribution from lightweight column statistics, extracts execution constraints from the target queries, and progressively adjusts the distribution to satisfy these constraints, ensuring the global distribution is maximally preserved while exact local cardinalities are strictly enforced. 

Extensive evaluations on the TPC-H and SSB benchmarks demonstrate that \dbrepro\ outperforms state-of-the-art baselines across comprehensive fidelity metrics. Compared to the data-driven baseline, it reduces cardinality error by up to 20.3\% while maintaining identical plan consistency. Against a workload-aware baseline, it reproduces 15\% more consistent execution plans and lowers the latency proportion error by 21.5\%. Furthermore, we validate its industrial practicality on a nearly 1\,TB real-world dataset from a specific industrial sector, managed by KingbaseES, successfully reproducing the execution performance of complex slow queries with high fidelity.
\end{abstract}



\maketitle

\input{1-intro}
\input{2-preliminary}

\input{3-Problem_formulation}

\input{4-methodology}
\input{5-evaluation}
\input{6-discussion}
\input{7-conclusion}
\input{8-acknowledgments}

\bibliographystyle{ACM-Reference-Format}
\balance
\bibliography{refers}

\end{document}

%% file: 1-intro.tex
\section{Introduction}

\begin{figure*}[!t]
    \centering 
    \includegraphics[width=0.98\textwidth]{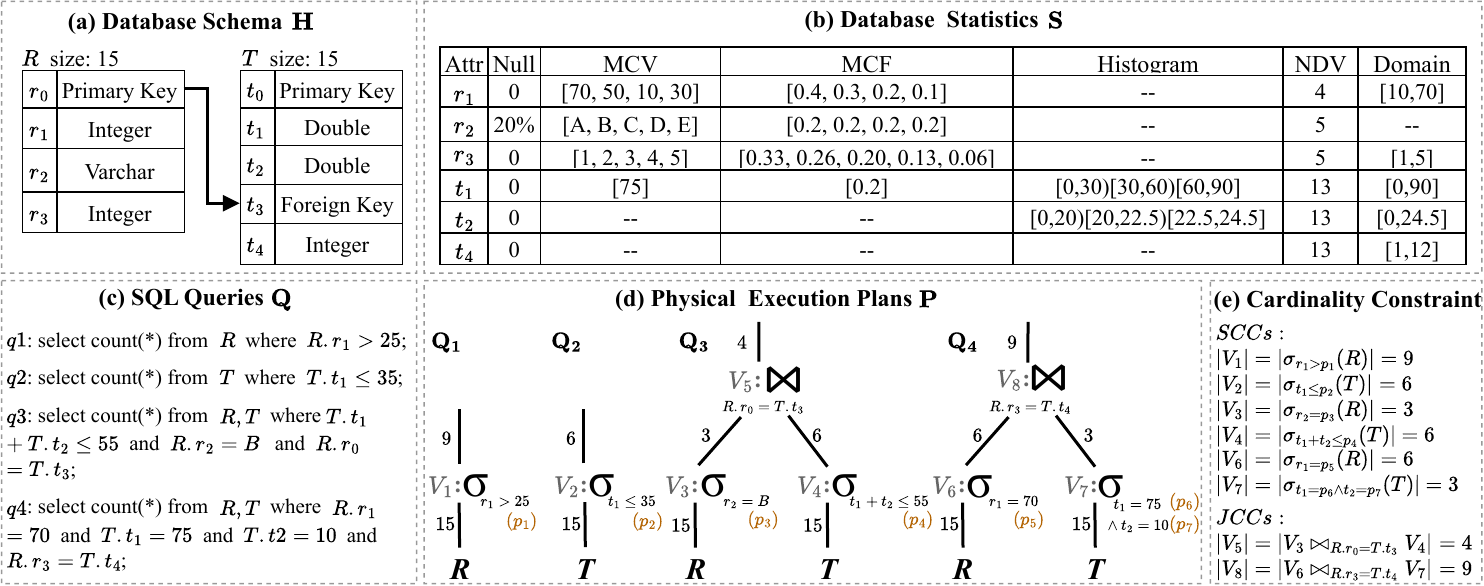} 
    \caption{Input example of database synthesis: database schema $H$ and statistics $S$, queries $Q$ and physical execution plans $P$.} 
    \Description{A multi-panel example with two relational tables, their primary-key and foreign-key schema, column statistics, four parameterized queries, and the corresponding physical execution plans annotated with operator cardinalities.}
    \label{fig:input_example}
\end{figure*}

In modern enterprise Database Management Systems (DBMSs), \textit{slow queries} represent a pervasive and critical challenge. Unlike lightweight transactional workloads, analytical queries typically involve complex multi-table joins, massive aggregations, and full-scale data scans. For example, in an enterprise data warehouse, a complex analytical query generating a large-scale operational report might suddenly degrade, taking hours instead of seconds to execute. Such performance anomalies monopolize critical CPU, memory, and I/O resources, causing severe contention that can precipitate a dramatic decline in overall system throughput or even service unavailability~\cite{10.14778/3389133.3389136}. Furthermore, they compromise the timeliness of data analytics, directly impeding an enterprise's capability for real-time decision-making.

When a slow query anomaly occurs, it must be promptly diagnosed and optimized (e.g., via index tuning or SQL rewriting). However, debugging the root causes of these anomalies directly on the live production instance is usually infeasible due to the severe risk of exacerbating online resource contention. Consequently, offline diagnosis has become the de facto approach. Yet, strict data privacy regulations prohibit DBAs from copying original user data to test environments. 
Therefore, synthesizing a proxy database via non-intrusive metadata is a critical step for diagnosing slow queries in an offline environment. In modern DBMSs, a SQL query is compiled by the query optimizer into a physical execution plan---a sequence of physical operations that dictates the query's performance. Consequently, the ultimate goal of database synthesis is to induce the query optimizer to yield physical execution plans identical to those in production. 
Evaluating a successful reproduction requires strict alignment in three dimensions: query plan structure, operator cardinalities, and execution times. 
Existing database synthesis techniques fall into two categories: \textit{data-driven} and \textit{workload-aware} approaches. \textit{Data-driven} methods~\cite{RSGen,10.1145/2723372.2735378,shadowDB,simple_realistic_data_gen,haritsa2015codd,soltana2017synthetic,buda2013cods,tay2013upsizer,buda2014vfds,ming2014bdgs,zhang2016dscaler,buda2017rex} reconstruct databases by learning column statistics or sampling production data. While they generally preserve the query plan structure, they fundamentally struggle to capture complex cross-column correlations and intricate multi-table join semantics. This limitation prevents them from enforcing strict local cardinalities. As a result, the simulated execution bottlenecks diverge significantly from the production environment, making it impossible to accurately preserve operator cardinalities and execution times. Conversely, \textit{workload-aware} methods~\cite{binnig2007qagen,veanes2010qex,lo2014mybenchmark,arasu2011data,sanghi2018hydra,sanghi2022projection,3277355.3277411,wang2024mirage,yang2022sam} formulate data generation as a constraint satisfaction problem derived from query plans. Although these methods enforce specific intermediate result sizes—thereby achieving strict cardinality consistency—they severely distort the global data distribution to satisfy these local constraints. Such distortion invariably misleads the optimizer's global cost model into selecting divergent physical execution plans. Ultimately, this approach fails to preserve the holistic query plan structure and the overall execution performance.

Achieving high-fidelity reproduction by satisfying both global statistics and local cardinalities presents three major challenges. First, reconciling the multivariate and heterogeneous inputs is difficult: database statistics dictate probabilistic data distributions, while workload cardinalities impose deterministic size constraints. Bridging this semantic gap to guide unified data generation remains an open problem. Second, reconciling potential conflicts between these inputs is non-trivial: global statistics are derived from database sampling, introducing approximation errors that may contradict the rigid intermediate cardinalities demanded by the execution plan. Resolving this conflict requires adjusting the data distributions without destroying the global statistical baseline. Third, slow queries in industrial workloads frequently involve non-PK-FK joins (e.g., on non-unique columns or composite keys). These joins trigger severe cardinality explosions—known as the fan-out effect—which significantly complicate constraint modeling and solving.


To overcome these challenges, we introduce \dbrepro, an automated end-to-end framework that formulates database synthesis as a constrained distribution synthesis problem, harmonizing the multivariate and heterogeneous inputs by synergistically integrating data-driven and workload-aware methods. 
Specifically, \dbrepro\ employs a novel hybrid constraint-solving approach to progressively adjust an initial statistics-based distribution, deriving a final distribution that satisfies all query constraints without compromising global characteristics. 
The hybrid constraint-solving approach adopts different constraint satisfiability solving methods for different types of constraints. It utilizes heuristic probabilistic inference combined with a distribution fusion algorithm to seamlessly integrate the Selection Cardinality Constraints (SCCs) into the global distribution, while resolving the potential conflicts. For Join Cardinality Constraints (JCCs), it tackles the fan-out problem in JCCs by embedding a \textit{Scaling Factor} mechanism into a Constraint Programming (CP) model, preventing cardinality explosions during non-PK-FK joins. 
In summary, our main contributions are as follows: 
\begin{itemize}[leftmargin=1em,topsep=0pt, label=$\triangleright$]
    \item We propose \dbrepro, an end-to-end automated framework that non-intrusively captures production telemetry and synthesizes test databases for high-fidelity slow query reproduction, enabling practical and reliable offline performance diagnostics.
    \item We design a novel hybrid constraint-solving approach to adjust the data distributions progressively. It utilizes heuristic inference and distribution fusion to reconcile statistical and cardinality conflicts, and introduces a Scaling Factor-enhanced constraint programming model to accurately resolve non-PK-FK joins. 
    \item We conduct extensive evaluations on the TPC-H and SSB benchmarks, as well as a nearly 1\,TB real-world industrial dataset. \dbrepro\ significantly outperforms state-of-the-art baselines by achieving superior high-fidelity reproduction and demonstrating robust practicality in a commercial DBMS.
\end{itemize}

%% file: 2-preliminary.tex
\section{Preliminaries}\label{section:Preliminaries}

Due to strict privacy regulations, data extracted from production databases is typically limited to aggregated statistics and workload artifacts. This section formalizes these non-intrusive inputs. Figure~\ref{fig:input_example} shows a minimal example with two tables, $R$ and $T$, and four queries, which we use throughout this paper.

\noindent \textbf{Database Schema ($H$)} defines tables, attributes, data types, primary keys (PKs), and foreign keys (FKs). In Figure~\ref{fig:input_example}a, $R$ has a PK $r_0$, and $T$ has an FK $t_3$ referencing $R.r_0$. The synthetic proxy database must satisfy these structural and referential integrity constraints.

\noindent \textbf{Database Statistics ($S$)} captures the global data distributions from the production catalog. In database systems, queries are executed according to physical execution plans compiled by the query optimizer. The prevailing Cost-Based Optimizers (CBOs) heavily rely on statistics as the foundational baseline to estimate execution costs and select the optimal physical operators and join orders~\cite{selinger1979access,chaudhuri1998overview,leis2015good}. They typically include: 
\begin{itemize}[leftmargin=*, label=$\triangleright$]
    \item \textit{Null Fraction:} The proportion of tuples containing NULL values.
    \item \textit{Number of Distinct Values (NDV):} The cardinality of an attribute's active domain.
    \item \textit{Most Common Values (MCVs) \& Frequencies (MCFs):} A set of the most frequent values coupled with their probabilities.
    \item \textit{Histograms:} A set of buckets (equi-depth or equi-width) summarizing the continuous data distribution.
\end{itemize}

\begin{figure*}[!t]
    \centering 
    \includegraphics[width=0.99\textwidth]{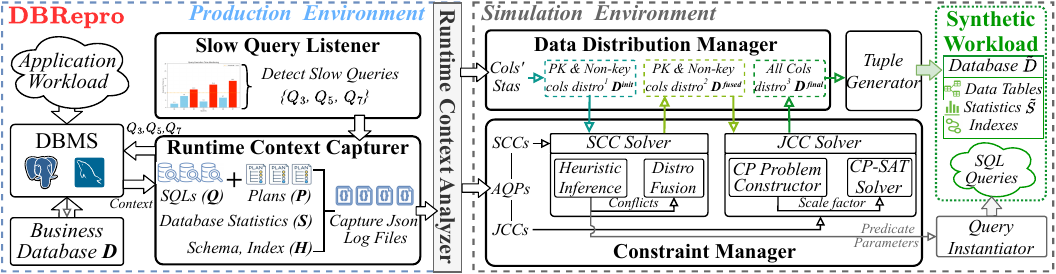} 
    \caption{Design overview of DBRepro.}
    \Description{The DBRepro workflow from production to an isolated simulation environment. A runtime context capturer collects schema, statistics, queries, and plans; an analyzer extracts constraints; distribution and constraint managers solve selection and join constraints; and a tuple generator materializes the proxy database.}
    \label{fig:design_overview}
\end{figure*}

In Figure~\ref{fig:input_example}b, $R.r_1$ has an NDV of 4 and an MCV set $\{70, 50, 10, 30\}$ with corresponding MCF $\{0.4, 0.3, 0.2, 0.1\}$, while $T.t_1$ uses three equi-depth histogram buckets ($[0,30)$, $[30,60)$, and $[60,90]$). These statistics serve as the foundational data distribution. Reproducing them ensures the optimizer selects the same join orders and physical operators, while distorting them produces divergent plans.

\vspace{0.5mm}
\noindent \textbf{Queries ($Q$)} denote a set of $n$ slow queries, $Q_1,$ $\ldots,$ $Q_n$, which are the original SQL statements executed in production.

\vspace{0.5mm}
\noindent \textbf{Physical Execution Plans ($P$)} are generated by the optimizer to dictate the exact sequence of physical operations for a query. Each query $Q_i$ is accompanied by a physical plan from the production environment, which inherently comprises three critical dimensions: query plan structure (join order and physical operator type), intermediate operator cardinalities, and execution time. Therefore, faithfully reproducing a slow query necessitates the precise reconstruction of these three elements in the simulated environment.


\vspace{0.5mm}
\noindent \textbf{Annotated Query Plans (AQPs $\hat{P}$)} are converted from the physical plans by parameterizing the constants in the predicates. 
For example, in Figure~\ref{fig:input_example}d, the predicate $t_1+t_2$ $\le 55$ in $Q_3$ is parameterized into $t_1+t_2$ $\le p_4$, and its output $V_4$ is labeled with a cardinality of 6. 
This parameterization significantly reduces the computational complexity. It avoids relying on computationally expensive models to synthesize exact tuples satisfying hardcoded constants~\cite{arasu2011data,sanghi2018hydra} and grants the solver the flexibility to straightforwardly infer new threshold parameters based on the generated data. 
The resulting query plan can precisely reconstruct the three elements: query plan structure, intermediate operator cardinalities, and execution time, required for slow query reproduction in the simulated environment.

\vspace{0.5mm}
\noindent \textbf{Cardinality Constraints (CCs)} are extracted from these AQPs to define the required output size of a physical operator. We represent the output of each query operator as a \textit{query operator view} ($V$). Based on $Q_3$ in Figure~\ref{fig:input_example}d, we derive the following CCs in Figure~\ref{fig:input_example}(e): $|V_3|: |\sigma_{r_2 = p_3}(R)| = 3$, $|V_4|: |\sigma_{t_1 + t_2 \le p_4}(T)| = 6$, and $|V_5|: |V_3 \bowtie_{R.r_0 = T.t_3} V_4| = 4$. We divide CCs into two categories:

\vspace{0.5mm}
\noindent \textbf{Selection Cardinality Constraints (SCCs)} are imposed by selection operators ($\sigma$) and introduce joint data distribution requirements among non-key columns. 
As shown in Figure~\ref{fig:input_example}(e), following \cite{wang2024mirage}, we categorize them into Unary Cardinality Constraints (UCCs) for single-column predicates (e.g., $V_3$), Arithmetic Cardinality Constraints (ACCs) for multi-column operations (e.g., $V_4$), and Logical Cardinality Constraints (LCCs) for predicates connected by logical operators (e.g., $V_7$). 


\vspace{0.5mm}
\noindent \textbf{Join Cardinality Constraints (JCCs)} are imposed by cross-table join operators ($\bowtie$), requiring specific joint distributions between the join keys. Figure~\ref{fig:input_example}(e) formulates the JCC for $V_8$ as $|V_6 \bowtie_{R.r_3 = T.t_4} V_7|=9$. This constraint requires that performing an equi-join on the specific keys ($R.r_3$ and $T.t_4$) between the two query operator views ($V_6$ and $V_7$) must yield exactly 9 matching rows.

%% file: 3-Problem_formulation.tex
\section{Problem Formulation}\label{section:ProblemFormulation}

In this section, we formalize the high-fidelity database synthesis problem for slow query reproduction. 
To achieve this, the synthesis process must strictly satisfy the cardinality constraints while maximally preserving the global statistical distributions defined by the database statistics. 

\begin{definition}[High-Fidelity Database Synthesis]\label{problem_define}
Given the target database schema $H$, statistics $S$, and a set of slow queries $Q$ with corresponding annotated query plans $\hat{P}$, the synthetic proxy database $\tilde{D}$ is expected to satisfy the following fidelity properties:
\begin{enumerate}[label=\arabic*), leftmargin=*]
    \item \textit{Schema Consistency:} $\tilde{D}$ strictly conforms to the structural and integrity constraints defined in $H$.
    \item \textit{Statistical Consistency:} The single-column statistics of $\tilde{D}$ (denoted $\tilde{S}$) closely match the production statistics ($\tilde{S} \approx S$), preserving the optimizer's global cost baseline.
    \item \textit{Plan Structural Consistency:} For each slow query $Q_i \in Q$, the generated physical plan $\tilde{P}_i$ is topologically identical to $\hat{P}_i$, preserving join orders and operator types.
    \item \textit{Cardinality Consistency:} 
    For each slow query $Q_i$, we evaluate the relative error between expected ($C_j$) and actual ($\tilde{C}_j$) outputs across its operators $O_{Q_i}$: $\text{RE}_{card}(Q_i) = \frac{\sum_{j \in O_{Q_i}} |C_j - \tilde{C}_j|}{\sum_{j \in O_{Q_i}} C_j}$.
    \item \textit{Latency Proportion Consistency:} 
    Directly comparing absolute execution times is unreliable due to potential differences in hardware and system configurations in the production and simulation environments. 
    Therefore, we evaluate the latency of each operator relative to the overall query latency to reliably reproduce the true performance bottlenecks. 
    For query $Q_i$, we evaluate the relative error between expected ($L_j$) and actual ($\tilde{L}_j$) latency proportions across key operators $K_{Q_i}$: $\text{RE}_{latency}(Q_i) = \frac{\sum_{j \in K_{Q_i}} |L_j - \tilde{L}_j|}{\sum_{j \in K_{Q_i}} L_j}$.
    \item \textit{Generalizability:} For unseen queries from the same application scenario, 
    $\tilde{D}$ continues to maintain Properties 3, 4, and 5.
\end{enumerate}
\end{definition}

%% file: 4-methodology.tex
\begin{figure*}[!t]
    \centering
    \includegraphics[width=\textwidth]{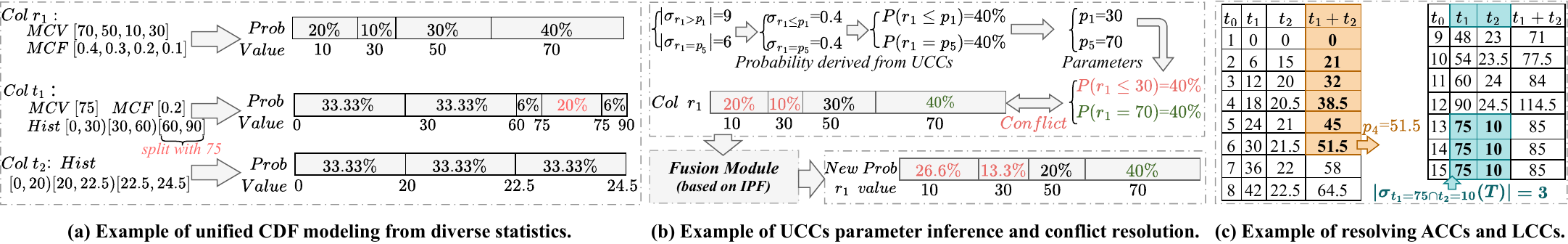}
    \caption{Examples of the statistical constraints modeling and selection cardinality constraints resolution algorithms.}
    \Description{Three examples illustrate the construction of a unified cumulative distribution from histograms and most-common values, parameter inference and distribution fusion for unary selection constraints, and sampling or tuple binding for arithmetic and logical constraints.}
    \label{fig:algorithms_overview}
\end{figure*}

\section{Methodology}

\subsection{Overview}

In this section, we present the design of \dbrepro. The core idea is to synthesize a high-fidelity proxy database that accurately reproduces the performance bottlenecks of slow queries without requiring access to sensitive production data. To achieve this, \dbrepro\ must handle multivariate inputs extracted from the production environment: catalog statistics and physical execution plans. In the isolated simulation environment, \dbrepro\ first uses the extracted statistics to model an initial global data distribution. It then employs a Hybrid Constraint-Solving Approach to systematically adjust this initial distribution, ensuring it strictly satisfies the exact cardinality constraints imposed by the queries. This refined final data distribution directly drives the synthetic proxy database generation.

Figure~\ref{fig:design_overview} shows the workflow. First, the \textit{Runtime Context Capturer} extracts runtime artifacts from the production servers: the database schema ($H$), column-level statistics ($S$), the queries ($Q$), and physical execution plans ($P$). The \textit{Analyzer} then transforms the raw physical plans into AQPs and extracts cardinality constraints. Based on these artifacts, the \textit{Data Distribution Manager} builds a global initial distribution (\textit{PK \& Non-key column distro$^1$ $\mathcal{D}^{init}$}) from the column statistics. Concurrently, the \textit{Constraint Manager} solves the SCCs and JCCs. To resolve conflicts between the cardinality constraints and this initial distribution, the \textit{SCC Solver} uses heuristic probabilistic inference combined with Iterative Proportional Fitting (IPF) to create a fusion distribution (\textit{PK \& Non-key column distro$^2$ $\mathcal{D}^{fused}$}). Then, the \textit{JCC Solver} uses this fusion distribution and the JCCs to construct a Constraint Programming (CP) model. For non-PK-FK joins, it uses a scaling factor-enhanced mechanism. Subsequently, the \textit{CP-SAT solver} solves this problem to obtain the foreign key distributions, which constitute the complete final distribution (\textit{All Cols distro$^3$ $\mathcal{D}^{final}$}). Finally, the \textit{Tuple Generator} uses this distribution to physically generate the synthetic proxy database.

\subsection{Runtime Context Capture and Analysis}\label{sec:capture}

\dbrepro\ begins with lightweight monitoring in the production environment, aiming to extract the necessary runtime context accurately while minimizing performance overhead. When the \textit{Slow Query Listener} detects slow queries, it triggers the \textit{Runtime Context Capturer}. Without accessing the underlying raw data, it extracts the necessary runtime context with minimal overhead. This context includes the database schema ($H$), column-level statistics ($S$), and the physical execution plans ($P$) annotated with operator cardinalities and execution latencies. Production DBMS catalogs (like \texttt{pg\_stats} in PostgreSQL) rely on periodic sampling and lack exact global aggregations like precise distinct counts or min/max bounds. For instance, min/max bounds can be accurately estimated from the outermost buckets of the equi-depth histograms, which typically contain hundreds of fine-grained intervals. When cardinality constraints involve these inferred bounds, the subsequent \textit{SCC Solver} strictly adjusts the data distribution to satisfy exact cardinalities, absorbing any initial boundary inaccuracies.

The \textit{Slow Query Listener} relies on native DBMS slow-query monitoring facilities. For example, PostgreSQL uses duration thresholds to log completed slow statements, while \texttt{auto\_explain}, when configured to log analyzed plans, records their physical execution plans. Upon detecting such records, \dbrepro\ performs post-execution collection: it obtains the logged queries and plans, identifies the referenced tables and columns, and retrieves their schema and column-level statistics from the system catalogs. This process neither re-executes the slow queries nor scans their base tables. Consequently, its cost is independent of the stored data volume and instead depends on the aggregate size of the recorded slow-query execution plans and the numbers of involved tables and columns, minimizing additional overhead on the production system.

The \textit{Runtime Context Analyzer} processes these extracted artifacts within the production environment. 
To simplify data generation, it applies relational algebra to push down selection operators to the bottom of the query tree~\cite{wang2024mirage}. This decouples the generation of non-key columns from PK-FK relationships, allowing \dbrepro\ to independently solve SCCs for non-key columns before resolving complex JCCs. It then transforms the raw physical plans ($P$) into AQPs by replacing hardcoded predicate constants with parameters (e.g., $r_1 > 35$ becomes $r_1 > p_1$). Finally, it extracts the cardinality constraints from these AQPs and categorizes them into formalized SCCs and JCCs. \dbrepro\ then transmits these processed artifacts to the simulation environment, routing the statistics ($S$) to the \textit{Data Distribution Manager} to initialize the data distribution, and the constraints to the \textit{Constraint Manager} for targeted solving.



\subsection{Distribution and Constraint Modeling}\label{sec:modeling}

In \dbrepro, database statistics establish the initial data distribution to anchor the optimizer's cost baseline, while query cardinalities serve as hard local constraints that dictate exact intermediate sizes to reproduce bottlenecks. To enable our hybrid constraint-solving algorithm, we must first model these heterogeneous inputs.


\noindent \textbf{Data Distribution Modeling.} Statistics usually contain discrete MCVs and continuous histograms. To enable heuristic probabilistic inference, we merge these structures into a Cumulative Distribution Function (CDF). If an MCV overlaps with a histogram bucket, we split the bucket at the MCV and proportionally allocate the remaining probability to the new sub-ranges based on their lengths (Figure~\ref{fig:algorithms_overview}a). For attributes with only MCVs ($r_1$) or histograms ($t_2$), the CDF directly maps values to probabilities. For attributes with both statistic types like $t_1$, the MCV $75$ ($20\%$ frequency) falls inside the bucket $[60, 90]$ ($33.3\%$ frequency). We split the bucket at $75$ and allocate the remaining $13.3\%$ evenly to the new sub-ranges $[60, 75)$ and $(75, 90]$ ($6.67\%$ each). To support efficient downstream tuple generation, we map this CDF into an array of memory-efficient buckets: $(low, high, frequency, nDistinct)$. This array serves as the initial data distribution for all PK and non-key columns, denoted as \textit{PK \& Non-key column distro$^1$ $\mathcal{D}^{init}$}, and is maintained by the \textit{Data Distribution Manager}. FK columns are excluded from this initialization because they are populated later to satisfy JCCs.


\begin{algorithm}[!t]
    \caption{Heuristic Inference and Distribution Fusion}
    \label{alg:param_instantiation}
    \DontPrintSemicolon
    \newcommand\mycommfont[1]{\small\itshape #1}
    \SetCommentSty{mycommfont}
    
    \KwIn{$\mathcal{C}$: UCCs with target probs $\{(op, p_i, v_{orig}, \tau)\}$; \\ \qquad \quad $\mathcal{D}^{init}$: \textit{Distro$^1$} (includes CDF $\mathcal{F}$ and MCV set $\mathcal{S}_{mcv}$)}
    \KwOut{$\mathcal{M}$: Parameter instantiations $\{p_i \to v\}$; \\ \quad \quad \quad \quad $\mathcal{D}^{fused}$: \textit{PK \& Non-key column distro$^2$}}
    
    $\mathcal{S}_{used} \gets \emptyset$, $\mathcal{M} \gets \emptyset$, $\mathcal{A} \gets \emptyset$\;
    Sort $\mathcal{C}$ in descending order of target probability $\tau$\;
    Merge constraints in $\mathcal{C}$ with $\Delta\tau \le \epsilon$ sharing parameters\;
    
    \ForEach{constraint $c = (op, p_i, v_{orig}, \tau) \in \mathcal{C}$}{
        \uIf(\tcp*[f]{Handle equality predicates via MCVs}){$op \text{ is } \text{`='}$}{
            $V_{cand} \gets \mathcal{S}_{mcv} \setminus \mathcal{S}_{used}$\;
            \eIf(\tcp*[f]{Find closest MCV frequency}){$V_{cand} \neq \emptyset$}{
                $v_{best} \gets \arg\min_{v \in V_{cand}} |freq(v) - \tau|$\;
                $\mathcal{S}_{used} \gets \mathcal{S}_{used} \cup \{v_{best}\}$\;
                $\mathcal{M}[p_i] \gets v_{best}$\;
                \If(\tcp*[f]{Record mismatch conflict}){$freq(v_{best}) \neq \tau$}{
                    $\mathcal{A}.\text{add}(\text{UPDATE\_MCV}(v_{best}, \tau))$\;
                }
            }(\tcp*[f]{Fallback to original constant}){
                $v_{orig} \gets \text{original query constant}$\;
                $\mathcal{S}_{used} \gets \mathcal{S}_{used} \cup \{v_{orig}\}$\;
                $\mathcal{M}[p_i] \gets v_{orig}$\;
                $\mathcal{A}.\text{add}(\text{ADD\_MCV}(v_{orig}, \tau))$\;
            }
        }
        \ElseIf(\tcp*[f]{Handle range predicates via CDF}){$op \in \{<, \le\}$}{
            $v_{pred} \gets \mathcal{F}^{-1}(\tau)$ \tcp*{Inverse CDF lookup}
            $\mathcal{M}[p_i] \gets v_{pred}$\;
            \If(\tcp*[f]{Record mismatch conflict}){$P(X \text{ op } v_{pred}) \neq \tau$}{
                $\mathcal{A}.\text{add}(\text{UPDATE\_RANGE}(v_{pred}, \tau, op))$\;
            }
        }
    }
    \eIf(\tcp*[f]{Apply IPF algorithm to reconcile conflicts}){$\mathcal{A} \neq \emptyset$}{
        $\mathcal{D}^{fused} \gets \text{IPF\_Distribution\_Fusion}(\mathcal{A}, \mathcal{D}^{init})$\;
    }{
        $\mathcal{D}^{fused} \gets \mathcal{D}^{init}$\;
    }
    \Return $\mathcal{M}, \mathcal{D}^{fused}$\;
\end{algorithm}

\noindent \textbf{Cardinality Constraints Modeling.}
The extracted SCCs and JCCs provide hard cardinality constraints in the form of absolute row counts (e.g., $|\sigma_{r_1 > p_1}(R)|$=$9$). However, the hybrid constraint solvers operate on probabilities. To bridge this gap, the \textit{Analyzer} normalizes these absolute cardinalities into relative target probabilities ($\tau$) by dividing them by the corresponding table sizes. For example, as shown in Figure~\ref{fig:algorithms_overview}b, the original constraint requires 6 rows. Since table $R$'s size is 15 (Figure~\ref{fig:input_example}a), we transform this into a parameterized probability constraint: $|\sigma_{r_1 \le p_1}(R)| = 0.4$ (using the complement $1 - \frac{9}{15}$). This transformation into a unified probabilistic domain enables the solver to perform heuristic probabilistic inference over the data distributions to strictly satisfy the required selectivity. 

\subsection{Hybrid Constraint Solving for DB Synthesis}\label{sec:solving}

This phase is the computational core of \dbrepro, bridging the \textit{Data Distribution Manager} and the \textit{Constraint Manager}. By systematically solving SCCs and JCCs, it progressively adjusts the distribution from the initial distribution \textit{$\mathcal{D}^{init}$} to the final distribution \textit{$\mathcal{D}^{final}$}. This adjustment harmoniously satisfies both the probabilistic global statistics and the deterministic workload cardinalities. To achieve this balance, we employ the \textit{Iterative Proportional Fitting (IPF)} algorithm. IPF is mathematically proven to converge on a new joint distribution that strictly satisfies new marginal constraints (the deterministic workload cardinalities) while minimizing the deviation (e.g., KL divergence) from the initial distribution (the probabilistic global statistics)~\cite{ipf_prove}.

\subsubsection{Solving Selection Cardinality Constraints} To accurately satisfy filter conditions without violating the underlying data distribution, we process SCCs progressively by complexity. We first resolve single-column UCCs through heuristic probabilistic inference and \textit{Distribution Fusion} mechanism, followed by satisfying LCCs via data binding and addressing ACCs through post-generation sampling.

\noindent \textbf{Resolving UCCs.} 
As shown in Algorithm~\ref{alg:param_instantiation}, the solver takes the initial probabilistic distribution ($\mathcal{D}^{init}$) and the extracted SCCs as inputs. It first normalizes greater-than predicates into less-than counterparts using the complement of their target probabilities ($1 - \tau$). It then sorts UCCs in descending order of target probabilities and merges constraints with nearly identical probabilities (within a threshold $\epsilon$) to share the same inferred parameter (Lines 2-3), thereby reducing solving overhead. 

The solver then employs operator-specific heuristic inference to satisfy these target probabilities. For equality predicates (Lines 5-17), the solver heuristically searches the MCV set for the best-matching frequency, using an exclusion set $\mathcal{S}_{used}$ to prevent mapping multiple parameters to the same MCV. If the MCV candidates are exhausted, it falls back to the original query constant. For range predicates (Lines 18-22), it performs an inverse lookup on the CDF $\mathcal{F}^{-1}(\tau)$ to locate the exact boundary threshold. If the chosen parameter's inherent statistical frequency does not perfectly match the required probability $\tau$, the solver records an adjustment action (e.g., \text{UPDATE\_MCV} or \text{UPDATE\_RANGE}) into an action list $\mathcal{A}$. The inferred parameters ($\mathcal{M}$) are routed to the \textit{Query Instantiator}, while the action list $\mathcal{A}$ is forwarded to \textit{Distribution Fusion}.

\noindent \textbf{Distribution Fusion.} 
When the action list $\mathcal{A}$ indicates conflicts between the query-required probabilities and the initial statistical distribution ($\mathcal{D}^{init}$), \textit{Distribution Fusion} intervenes (Lines 23-24). While database statistics establish the optimizer's foundational cost baseline, they are inherently approximate due to periodic sampling. Therefore, instead of treating them as rigid, immutable constraints, we strategically adjust these global distributions only when necessary---specifically, when they conflict with the exact marginal constraints dictated by the SCCs ($\mathcal{A} \neq \emptyset$). This ensures that the synthetic proxy database exactly reproduces local execution bottlenecks. Guided by the IPF algorithm, we adjust the specific attribute distributions to strictly satisfy the new marginal constraints while mathematically minimizing the deviation from $\mathcal{D}^{init}$. As illustrated in Figure~\ref{fig:algorithms_overview}b, suppose the solver receives two parameterized SCCs requiring $P(r_1 \le p_1) = 40\%$ and $P(r_1 = p_5) = 40\%$. By searching $\mathcal{D}^{init}$, it infers $p_1$ = $30$ and $p_5$ = $70$. However, the initial distribution indicates $P(r_1 \le 30)$ is only $30\%$, triggering a conflict. Through IPF, the distribution of $r_1$ is adjusted to $\{26.67\%, 13.33\%, 20\%, 40\%\}$. This new fusion distribution---denoted as \textit{PK \& Non-key column distribution$^2$ $\mathcal{D}^{fused}$}---strictly satisfies the local $40\%$ cardinality requirement while preserving the global relative proportions.


\begin{figure}[!t]
    \centering 
    \includegraphics[width=\linewidth]{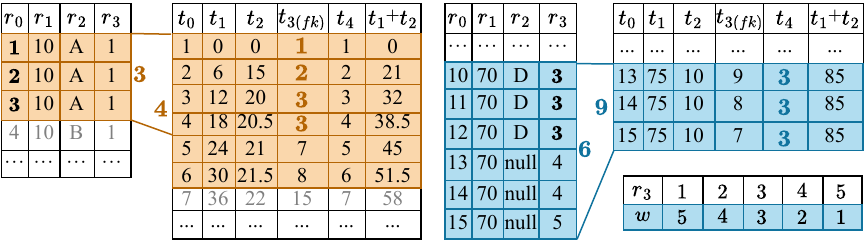} 
    \begin{subfigure}[t]{0.43\linewidth}
        \caption{PK-FK JCC.}
        \label{fig:solve_pkfk_jcc}
    \end{subfigure}
    \hfill
    \begin{subfigure}[t]{0.52\linewidth}
        \caption{Non-PK-FK JCC.}
        \label{fig:solve_nonkey_jcc}
    \end{subfigure}
    \caption{Resolution of Join Cardinality Constraints (JCCs).}
    \Description{Two examples compare join-cardinality resolution. The left assigns foreign keys for a primary-key to foreign-key join, while the right uses key frequencies as scaling factors to control fan-out in a non-primary-key join.}
    \label{fig:solve_jccs_overall}
\end{figure}

\noindent \textbf{Resolving ACCs and LCCs.} As shown in Figure~\ref{fig:algorithms_overview}c, arithmetic constraints (e.g., $|\sigma_{t_1 + t_2 \le p_4}(T)| = 6$) and logical constraints (e.g., $|\sigma_{t_1 = 75 \land t_2 = 10}(T)| = 3$) involve multi-column dependencies that cannot be directly inferred from 1D marginal distributions. Building upon the decoupling principles established in prior work~\cite{wang2024mirage}, we apply a post-generation approach to resolve these dependencies based on $\mathcal{D}^{fused}$. For ACCs (left part of Figure~\ref{fig:algorithms_overview}c), we sample a batch of these generated tuples to compute their arithmetic expressions, ensuring a bounded approximation error~\cite{hoeffding1963probability}. Evaluating these samples allows us to accurately infer the required parameters (e.g., $p_4=51.5$) to fulfill the ACCs. For LCCs (right part of Figure~\ref{fig:algorithms_overview}c), the \textit{Tuple Generator} physically generates the non-key columns, explicitly binding values required by LCCs to the same rows to guarantee their co-occurrence.

\subsubsection{Solving Join Cardinality Constraints} 
After deriving $\mathcal{D}^{fused}$, \dbrepro must resolve the inter-table cardinality dependencies imposed by JCCs to derive the complete final distribution (\textit{All Cols distro$^3$} $\mathcal{D}^{final}$). This requires accurately populating foreign keys.

\noindent \textbf{Resolving PK-FK JCCs.} 
For standard PK-FK joins, existing methods~\cite{wang2024mirage} formalize foreign key population based on a known set of primary keys. Given a join operation between the left view $V_l$ (typically the build side) and the right view $V_r$ (the probe side) with a target cardinality $n_{jcc}$, two fundamental \textit{P}opulating \textit{F}K rules ($PF$) are derived. Rule (1) ensures that selecting PKs from $V_l$ to populate FKs in $V_r$ yields exactly $n_{jcc}$ matched rows. Rule (2) guarantees that the remaining rows in $V_r$ are populated using PKs absent from $V_l$ to prevent unintended join outputs:
\begin{equation*}
\begin{aligned}
\text{(1)}~&PF_{V_l \rightarrow V_r} = n_{jcc},\\
\text{(2)}~&PF_{\overline{V_l} \rightarrow V_r} = |V_r| - n_{jcc}.
\end{aligned}
\end{equation*}
In Figure~\ref{fig:solve_jccs_overall}a (representing the PK-FK join in $Q_3$), the target cardinality is 4. $V_l$ provides 3 available PKs (highlighted in yellow), and $V_r$ has 6 rows satisfying the ACCs. Because $R.r_0$ is a unique primary key, we select 3 PKs (e.g., 1, 2, 3) from $V_l$ to populate 4 FKs in $V_r$, and fill the remaining 2 FKs with absent keys (e.g., 7, 8). This allocation perfectly meets the target cardinality of 4.

\noindent \textbf{Resolving Non-PK-FK JCCs.} In industrial scenarios, queries frequently involve joins on a subset of composite primary keys or between two non-key columns. This violates the stringent prerequisite of PK uniqueness required by traditional $PF$ rules. If the join column in $V_l$ contains duplicate values, treating it as a unique key and blindly applying the standard $PF$ rules inevitably triggers a Cartesian product explosion (fan-out effect). The physical cardinality will diverge unpredictably, misleading the execution plan.

To overcome this limitation, we introduce a \textbf{Scaling Factor} ($w$) mechanism. Unlike existing solvers that strictly assume primary key uniqueness and inevitably fail on complex analytical joins, our scaling factor explicitly models the empirical fan-out effect. Specifically, \dbrepro\ builds a pseudo-PK column for $V_l$ based on its distribution and uses the frequency of each key value $k$ within $V_l$ as its scaling factor $w_k$. This transforms the rigid $PF$ rules into weighted constraint equations:
\begin{equation*}
\begin{aligned}
\text{(3)}~&\sum_{k \in V_l} (w_k \times PF_{k \rightarrow V_r}) = n_{jcc},\\
\text{(4)}~&\sum_{k \notin V_l} PF_{k \rightarrow V_r}
= |V_r| - \sum_{k \in V_l} PF_{k \rightarrow V_r}.
\end{aligned}
\end{equation*}
where $PF_{k \rightarrow V_r}$ is the number of times key $k$ is populated into the right view $V_r$'s FK column. 
Figure~\ref{fig:solve_jccs_overall}b shows the non-PK-FK join $r_3$ = $t_4$ in $Q_4$, where $n_{jcc}$=$9$. In Figure~\ref{fig:input_example}d, $V_r$ has 3 rows satisfying the LCC. While the base column $r_3$ contains duplicates, the filtered view $V_l$ retains only keys $\{3,4,5\}$ with local frequencies $w$=$\{3,2,1\}$ (bottom-right in Figure 4b). 
Previous algorithms would fail here: if they blindly selected distinct values from $r_3$ to populate $t_4$, causing the join output to explode due to duplicates. Instead, \dbrepro\ uses the scaling factors $w_k$ to form the weighted equation: $3 \cdot PF_{3 \to t_4} + 2 \cdot PF_{4 \to t_4} + 1 \cdot PF_{5 \to t_4} = 9
$. The solver can easily find a feasible solution, like selecting key $3$ to populate all 3 rows in $V_r$ yields exactly $3 \times 3 = 9$ matches. If $V_7$ had more rows that need to be mismatched, they would be safely populated with absent keys from column $r_3$ (e.g., $k \notin V_6$, like $k=1,2$) to strictly meet the cardinality without explosion.

When queries contain multiple join constraints across several tables, the $PF$ rules and weighted equations combine into a comprehensive Constraint Programming (CP) model. This allows \dbrepro\ to solve both strict PK-FK and non-PK-FK joins within a unified framework using CP-SAT solvers like OR-Tools~\cite{perron2021ortools}. For acceleration when absolute precision is less critical, the system can dynamically estimate a uniform scaling factor using either sampling methods or the \textit{n\_distinct} statistic to approximate the degree of uniqueness. Ultimately, we compute the valid FK distributions, which completes the evolution into the final $\mathcal{D}^{final}$.

Through the systematic resolution of SCCs and JCCs, \dbrepro\ computes the valid data distributions for all attributes. This completes the entire data distribution evolution from the initial statistics-based $\mathcal{D}^{init}$ to the final constraint-compliant $\mathcal{D}^{final}$. Through this progressive evolution, \dbrepro\ satisfies both the probabilistic global statistics and the deterministic local cardinalities. The resulting $\mathcal{D}^{final}$ serves as the definitive blueprint, directly driving the \textit{Tuple Generator} to physically synthesize the high-fidelity proxy database required for reproducing queries.

%% file: 5-evaluation.tex
\section{Evaluation}

We structure our evaluation around three research questions:

\noindent \textbf{RQ1:} How does \dbrepro compare to existing generation methods in high-fidelity slow query reproduction and end-to-end efficiency?

\noindent \textbf{RQ2:} How well does \dbrepro generalize to unseen queries on the same benchmark schema?

\noindent \textbf{RQ3:} How does \dbrepro\ perform on commercial DBMSs in real-world industrial scenarios?

\subsection{Experimental Setup}

\subsubsection{Environment and DBMS}

For RQ1 and RQ2, we ran experiments on a server with an Intel Xeon Gold 5220 CPU @ 2.20\,GHz (36 cores) and 156\,GB RAM using PostgreSQL 14.11. Since the cardinality-constraint formulations are DBMS-independent, PostgreSQL provides a widely used open-source CBO platform for these experiments. For the industrial-scale practicality evaluation in RQ3, we deployed \dbrepro\ on a server with dual HiSilicon Kunpeng-920 processors (128 cores) and 1\,TB RAM using KingbaseES V8R6C9B14, a proprietary commercial RDBMS managing the terabyte-scale data. Together, the two systems represent open-source and commercial CBO-based DBMSs.





\subsubsection{Baseline Methods}
We compare \dbrepro\ against state-of-the-art data-driven and workload-aware baseline methods:
\begin{itemize}
    \item \rsgen~\cite{RSGen}: A data-driven generator that reverses column-level statistics to reproduce the original data distribution. It is entirely workload-independent and does not enforce any query-specific cardinality constraints.
    \item \hydra~\cite{sanghi2018hydra}: A workload-aware regenerator that models plan cardinalities through linear programming. It requires schema anonymization at the client side, which causes semantic loss, and its LP formulation supports only a restricted set of filter and equi-join operators.
    \item \touchstone~\cite{3277355.3277411}: A workload-aware generator that decomposes query plan trees into constraint chains to satisfy cardinality constraints and support scalable parallel generation. It does not support complex logical predicates.
    \item \mirage~\cite{wang2024mirage}: A recent state-of-the-art workload-aware approach tailored for complex analytical workloads. It strictly bounds cardinality deviations through a multi-phase generation strategy and offers the broadest operator support among existing workload-aware methods.
\end{itemize}

\subsection{RQ1: Reproduction Fidelity}

To answer RQ1, we evaluate the end-to-end reproduction fidelity of \dbrepro\ against baselines on the TPC-H~\cite{tpch} and SSB~\cite{10.1007/978-3-642-10424-4_17} benchmarks. We use scale factor \textit{SF=1} to assess reproduction fidelity and \textit{SF=200} to evaluate synthesis efficiency. The synthesis quality is measured based on the properties defined in Definition~\ref{problem_define}. Specifically, we report the \textit{plan structural match rate} (Property 3), the mean relative error of cardinality ($\text{RE}_{card}$, Property 4), and the mean relative error of latency proportion ($\text{RE}_{latency}$, Property 5). We also measure the end-to-end efficiency via metadata extraction and data generation time. As detailed below, \dbrepro\ achieves the strongest overall fidelity across all metrics among the tested methods.


\begin{table}[!t]
    \centering
    \caption{Plan structural consistency on the TPC-H and SSB. The first number in the \textbf{Consistent} column accumulates strictly identical and topologically consistent cases, with the parenthetical value being the topologically consistent ones.  
    The \textbf{Inconsistent} column tallies category (iii).}
    \label{tab:plan_consistency_detail}
    \vspace{-3mm}
    \setlength{\tabcolsep}{3pt} 
     \resizebox{\linewidth}{!}{
        \begin{tabular}{lcccc}
            \toprule
            \textbf{Generation} & \multicolumn{2}{c}{\textbf{TPC-H (22 Queries)}} & \multicolumn{2}{c}{\textbf{SSB (13 Queries)}} \\
            \cmidrule(lr){2-3} \cmidrule(lr){4-5}
            \textbf{Methods} & \textbf{Consistent} & \textbf{Inconsistent} & \textbf{Consistent} & \textbf{Inconsistent} \\
            \midrule
            \hydra           & 3 (0) & 19 & 7 (4) & 6 \\
            \touchstone & 9 (1) & 13 & 10 (8) & 3 \\
            \mirage        & 15 (6) & 7 & 12 (3) & 1 \\
            \rsgen         & 18 (4) & 4 & 13 (4) & 0 \\
            \dbrepro                  & \textbf{18 (4)} & \textbf{4} & \textbf{13 (4)} & \textbf{0} \\
            \bottomrule
        \end{tabular}
    } 
\end{table}

\subsubsection{Plan Structural Consistency}\label{eval:planconsistency}

To provide a granular analysis of structural fidelity, we classify plan consistency into three categories: (i) \textit{Strictly Identical}: the synthetic plan perfectly matches the production plan in join order and physical operator types; (ii) \textit{Topologically Consistent}: the join order matches, but some physical operator types differ (e.g., hash join vs.\ merge join); and (iii) \textit{Inconsistent}: join orders diverge, or the baseline fails to support the query's operators, rendering subsequent operator-level comparisons invalid. 

As shown in Table~\ref{tab:plan_consistency_detail}, \dbrepro\ and the data-driven method \rsgen\ achieve the highest structural consistency. The workload-aware methods ignore global statistics and distort the cost model's baseline by enforcing cardinality constraints, which mislead the optimizer into selecting divergent join orders. Conversely, data-driven methods preserve these macroscopic distributions. \dbrepro\ harmonizes these paradigms by solving local constraints within the bounds of the global statistical baseline, maintaining structural consistency.

\subsubsection{Relative Error of Cardinality}\label{sec:cardinality}

To comprehensively evaluate downstream fidelity across both the cardinality analysis in this section and the latency evaluation in Section~\ref{sec:latency}, we employ two distinct metrics inspired by prior works~\cite{wang2024mirage,huang2025query}: (1) \textit{Valid-Queries Mean}, which calculates the error exclusively for queries with consistent execution plans to assess the underlying constraint satisfaction accuracy; and (2) \textit{All-Queries Mean}, which evaluates the overall effectiveness across the benchmark by assigning a 100\% error to structurally inconsistent or unsupported queries.

\begin{figure}[!t] 
    \centering
        \includegraphics[height=0.55cm]{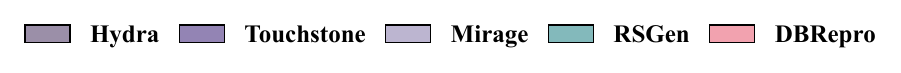}
        \vspace{-1mm} 
        
        \begin{subfigure}[b]{0.32\linewidth}
            \centering
            \includegraphics[width=\linewidth]{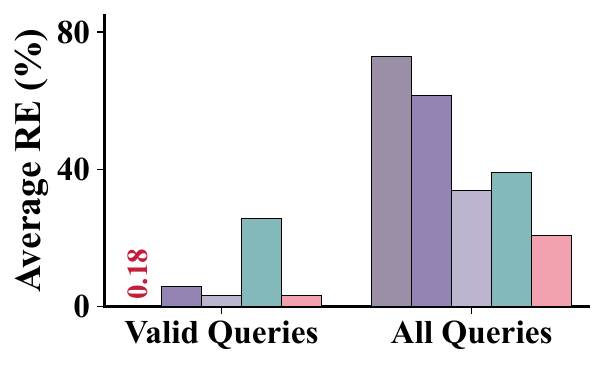}
            \vspace{-6mm}
            \caption{Avg SCCs $\text{RE}_{card}$}
            \label{fig:scc_tpch}
        \end{subfigure}
        \hfill
        \begin{subfigure}[b]{0.32\linewidth}
            \centering
            \includegraphics[width=\linewidth]{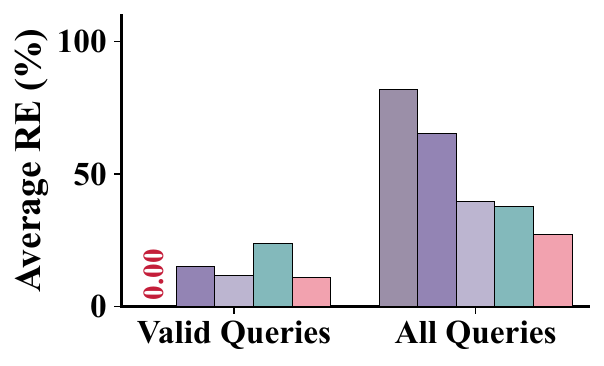}
            \vspace{-6mm}
            \caption{Avg JCCs $\text{RE}_{card}$}
            \label{fig:jcc_tpch}
        \end{subfigure}
        \hfill
        \begin{subfigure}[b]{0.32\linewidth}
            \centering
            \includegraphics[width=\linewidth]{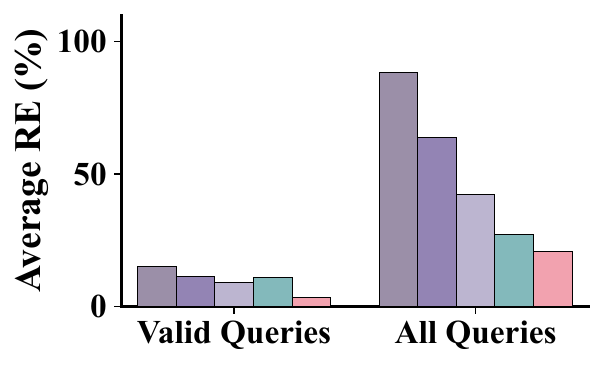}
            \vspace{-6mm}
            \caption{Avg $\text{RE}_{latency}$}
            \label{fig:latency_tpch}
        \end{subfigure}
        
        \vspace{1mm} 
        
        \begin{subfigure}[b]{0.32\linewidth}
            \centering
            \includegraphics[width=\linewidth]{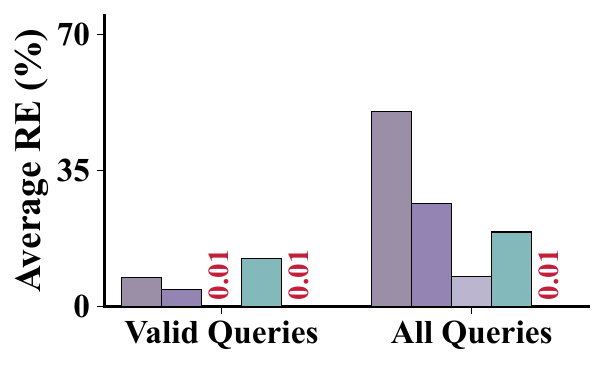}
            \vspace{-6mm}
            \caption{Avg SCCs $\text{RE}_{card}$}
            \label{fig:scc_ssb}
        \end{subfigure}
        \hfill
        \begin{subfigure}[b]{0.32\linewidth}
            \centering
            \includegraphics[width=\linewidth]{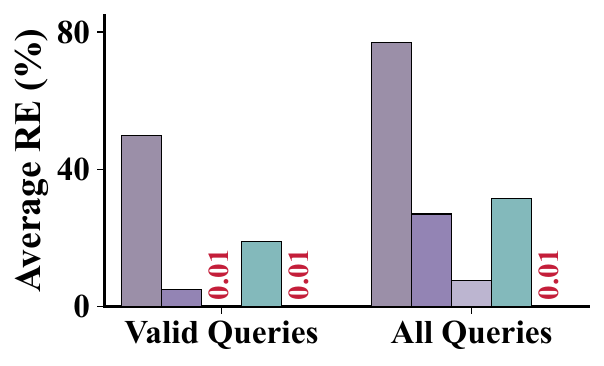}
            \vspace{-6mm}
            \caption{Avg JCCs $\text{RE}_{card}$}
            \label{fig:jcc_ssb}
        \end{subfigure}
        \hfill
        \begin{subfigure}[b]{0.32\linewidth}
            \centering
            \includegraphics[width=\linewidth]{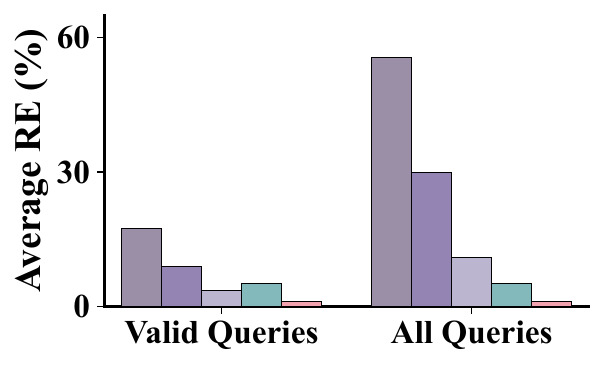}
            \vspace{-6mm}
            \caption{Avg $\text{RE}_{latency}$}
            \label{fig:latency_ssb}
        \end{subfigure}
    
        \caption{Comparison of average relative error of cardinality ($\text{RE}_{card}$) for SCCs and JCCs, and latency proportion ($\text{RE}_{latency}$) on TPC-H (top row) and SSB (bottom row).}
        \Description{Six grouped bar charts compare Hydra, Touchstone, Mirage, RSGen, DBRepro, and PostgreSQL where applicable. The top row reports selection-cardinality, join-cardinality, and latency-proportion errors for TPC-H, and the bottom row reports the same three metrics for SSB. DBRepro has the lowest or near-lowest errors across the comparisons.}
        \label{fig:performance_error_overall}
\end{figure}

Figure~\ref{fig:performance_error_overall} shows the average relative cardinality error ($\text{RE}_{card}$) for selection and join operators. Data-driven methods (\rsgen) assume column independence and thus fail to capture complex multi-column or cross-table correlations. This causes intermediate cardinalities to exhibit unbounded growth, averaging a 12.4\% error for selection operators and exceeding 25.6\% for join operators on \textit{valid-queries}. Workload-aware methods (\hydra, \touchstone), conversely, successfully achieve lower \textit{valid-queries} errors by explicitly modeling intermediate cardinality constraints. However, their performance severely deteriorates on \textit{all-queries} due to two limitations. First, mathematically modeling diverse and complex operators is inherently challenging, strictly limiting their applicability to a narrow subset of queries. Second, as shown in Section~\ref{eval:planconsistency}, their distorted global distributions induce inconsistent plans. Consequently, their successfully satisfied local constraints become entirely ineffective during actual execution, increasing their \textit{all-queries} mean error.

By driving data generation through progressive distribution evolution, \dbrepro\ explicitly models exact cardinality constraints while preserving the vast majority of non-constrained data in line with authentic statistics. This dual-guarantee allows \dbrepro\ to achieve a remarkably low cardinality error on \textit{valid-queries}, while delivering the lowest \textit{all-queries} mean error across all benchmarks, reducing the error by 20.3\% compared to data-driven baselines.

\subsubsection{Relative Error of Latency Proportion}\label{sec:latency}

Achieving latency fidelity (Figure~\ref{fig:performance_error_overall} (c) and Figure~\ref{fig:performance_error_overall} (f)) is drastically more complex than satisfying cardinalities, as execution time is sensitive to physical overheads like data skew, CPU hash collisions, and I/O spilling.

Workload-aware methods struggle with latency reproduction due to two inherent limitations. First, by ignoring global statistics, they disrupt the Cost-Based Optimization (CBO) mechanism~\cite{leis2015good}. This misleads the optimizer into selecting divergent physical operators (e.g., swapping a Hash Join for a Nested Loop Join), altering execution time complexity. Second, methods like \hydra\ and \touchstone\ neglect data distribution metrics such as NDV, inducing data skew and thus triggering severe hash collisions and unintended disk spilling. Data-driven methods (\rsgen) successfully preserve operator types, enabling a low 5\% latency proportion error on the simpler SSB. However, on complex workloads like TPC-H, their inability to enforce cross-column constraints triggers cascading intermediate data explosions. This overestimation increases memory consumption and physical costs for downstream operators, raising their comprehensive \textit{all-queries} latency proportion error to 27\%.

\dbrepro\ addresses the limitations of both approaches. By providing the optimizer with accurate statistics, it preserves operator types and plan structures. Concurrently, its exact cardinality control prevents unintended memory spilling and physical overheads. As a result, \dbrepro\ achieves the lowest mean latency proportion error on \textit{valid-queries} (3.29\% on TPC-H and 1.06\% on SSB), reducing the error by 21.5\% compared to workload-aware methods.

\begin{figure}[!t]
    \centering
    \includegraphics[width=0.75\linewidth]{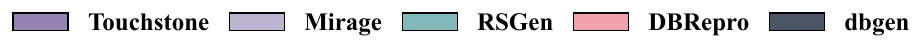}
    
    \begin{subfigure}[b]{0.49\linewidth}
        \centering
        \includegraphics[width=\linewidth]{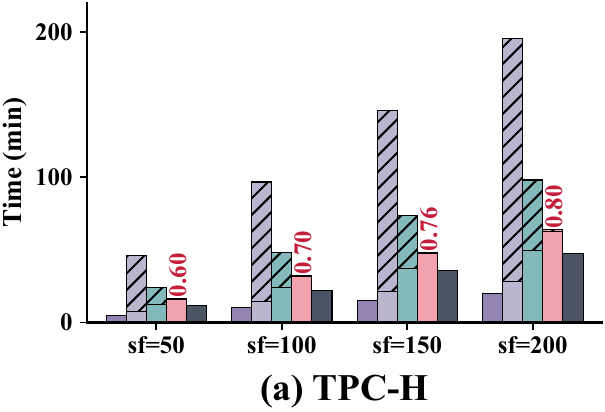}
        \label{fig:total_time_tpch}
    \end{subfigure}\hfill
    \begin{subfigure}[b]{0.49\linewidth}
        \centering
        \includegraphics[width=\linewidth]{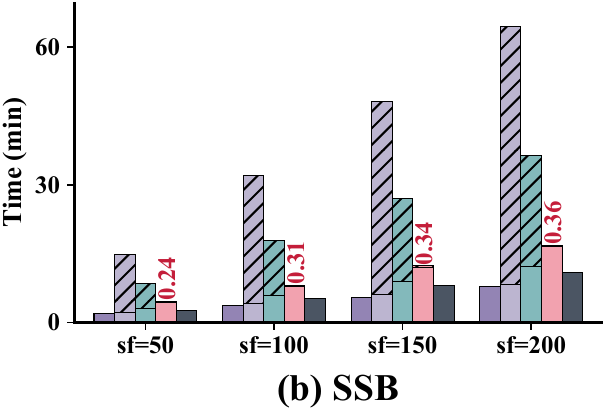}
        \label{fig:total_time_ssb}
    \end{subfigure}
    \vspace{-6mm}
    \caption{Overall extraction (upper hatched) and generation (lower solid) efficiency. }
    \Description{Two stacked bar charts compare metadata extraction time and database generation time for the evaluated methods on TPC-H and SSB. Hatched upper segments denote extraction and solid lower segments denote generation.}
    \label{fig:total_time_overall}
    \vspace{-4mm}
\end{figure}

\subsubsection{Extraction and Generation Efficiency}\label{sec:efficiency}

To evaluate end-to-end efficiency, we measure the total execution time across varying scale factors and decompose it into online context extraction and offline data generation. As shown in Figure~\ref{fig:total_time_overall}, the hatched and solid bars represent these two stages, respectively. \touchstone\ reports only its generation time because the constraint formulation must be manually constructed from the target workload. This human effort does not constitute an automated extraction stage and is not included in its reported runtime; consequently, its end-to-end cost is not directly comparable with those of automated methods. In contrast, \textit{dbgen} generates standard benchmark data directly from predefined specifications and therefore requires no production-context extraction.

During data generation, \dbrepro\ is approximately $2\times$ slower than \mirage. This is an anticipated trade-off: \dbrepro's hybrid algorithm must initially synthesize baseline data to satisfy global statistics before constructing and solving the CP problem for JCCs. However, because this generation occurs entirely offline in the simulation environment, it is highly insensitive to time overhead. Conversely, the extraction phase executes directly on the resource-critical production database, making the minimization of intrusion paramount. As illustrated, \dbrepro\ extracts existing slow-query execution plans and catalog metadata without full-table scans. Its extraction time is therefore independent of the underlying data volume and instead depends on the aggregate size of the recorded slow-query execution plans and the numbers of their referenced tables and columns. In contrast, \rsgen\ and \mirage\ require expensive full-table aggregation queries to derive precise global bounds, causing their extraction times to scale linearly with data volume. This reduction in online extraction overhead compensates for the offline generation cost, allowing \dbrepro\ to achieve roughly $1.5\times$ and $3\times$ the overall efficiency of \rsgen\ and \mirage\, respectively.

\subsection{RQ2: Generalization and Statistical Fidelity}

To answer RQ2, we conduct two interconnected evaluations. First, we test \textit{generalizability} using 20 unseen queries (10 per benchmark, spanning three complexity levels) synthesized by \textit{GPT-4o}. The queries were generated from the benchmark schemas and workloads and manually verified for syntactic correctness and successful execution. Second, to explain the statistical basis of this generalization capability, we assess \textit{statistical consistency} by directly measuring the recovery accuracy of the underlying database statistics and data densities across value ranges.


\subsubsection{Performance on LLM-Synthesized Unseen Workloads}

We compare the reproduction fidelity of proxy databases generated by each method. \hydra\ is excluded because its forced schema anonymization and redundant data generation make this generalization test inapplicable.

\begin{table}[!t]
    \centering
    \caption{Plan structural consistency on LLM-synthesized unseen queries across different complexity levels.}
    \label{tab:unseen_queries_consistency}
    \vspace{-3mm}
    \setlength{\tabcolsep}{3.5pt} 
    \resizebox{\linewidth}{!}{
        \begin{tabular}{lcccccc}
            \toprule
            \textbf{Generation} & \multicolumn{3}{c}{\textbf{TPC-H (10 Queries)}} & \multicolumn{3}{c}{\textbf{SSB (10 Queries)}} \\
            \cmidrule(lr){2-4} \cmidrule(lr){5-7}
            \textbf{Methods} & \textbf{Low(3)} & \textbf{Med(4)} & \textbf{High(3)} & \textbf{Low(3)} & \textbf{Med(4)} & \textbf{High(3)} \\
            \midrule
            \touchstone & 2 & 1 & 0 & 3 & 2 & 0 \\
            \mirage     & 3 & 2 & 0 & 3 & 3 & 0 \\
            \rsgen      & 3 & 4 & 1 & 3 & 4 & 2 \\
            \dbrepro    & \textbf{3} & \textbf{4} & \textbf{1} & \textbf{3} & \textbf{4} & \textbf{3} \\
            \bottomrule
        \end{tabular}
    }
\end{table}

\begin{figure}[!t]
    \centering
    \includegraphics[height=0.55cm]{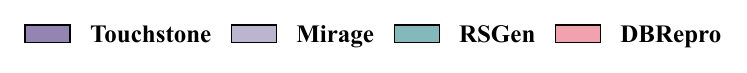}
    \vspace{-2mm}
    
    \begin{subfigure}[b]{0.48\linewidth}
        \centering
        \includegraphics[width=\linewidth]{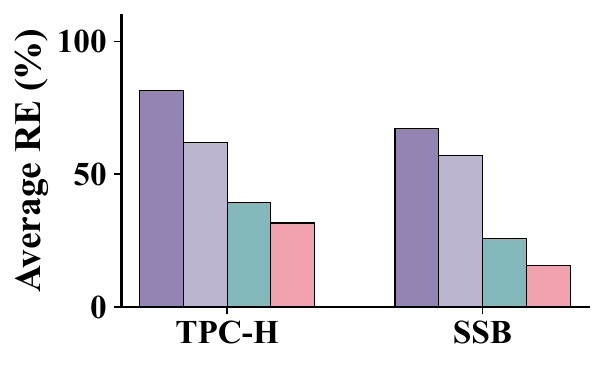}
        \caption{Average $\text{RE}_{card}$}
        \vspace{-2mm}
        \label{fig:unseen_card_err}
    \end{subfigure}
    \hfill 
    \begin{subfigure}[b]{0.48\linewidth}
        \centering
        \includegraphics[width=\linewidth]{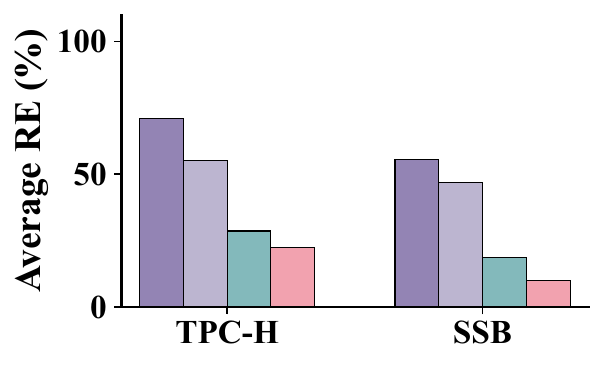}
        \caption{Average $\text{RE}_{latency}$}
        \vspace{-2mm}
        \label{fig:unseen_latency_err}
    \end{subfigure}
    \caption{Average relative error of cardinality and latency proportion on LLM-synthesized unseen queries.}
    \Description{Two grouped bar charts compare Touchstone, Mirage, RSGen, and DBRepro on twenty unseen queries. The left chart reports average cardinality error and the right reports average latency-proportion error for TPC-H and SSB; DBRepro records the lowest overall errors.}
    \label{fig:unseen_queries_performance}
\end{figure}

As shown in Table~\ref{tab:unseen_queries_consistency} and Figure~\ref{fig:unseen_queries_performance}, workload-aware methods struggle with high-complexity unseen queries. Novel queries introduce different predicate combinations and join topologies; without a global distribution foundation, the optimizer is easily driven toward divergent plans, substantially increasing their \textit{all-queries} errors. In contrast, both \rsgen\ and \dbrepro\ preserve macroscopic execution structures across the three complexity levels, with \dbrepro\ improving overall fidelity by approximately 10\% over \rsgen.

\rsgen's inability to model multi-column distributions can cause cascading cardinality underestimation on unseen queries with cross-column predicates. \dbrepro's advantage stems from two factors. First, the historical workloads used during synthesis cover diverse business scenarios, improving estimates for unseen queries that share similar predicates or join keys. Second, satisfying complex constraint chains during generation implicitly embeds business semantics and joint distributions into the physical data. Consequently, the optimizer encounters more faithful physical correlations even for complex unseen queries, supporting \dbrepro's generalization beyond the synthesis workload.



\subsubsection{Fidelity on Synthesized Databases' Statistics}

To examine the statistical basis of \dbrepro's generalization and structural consistency, we evaluate the fidelity of the underlying data distributions. We measure relative errors in three core statistical components extracted from each proxy database: NDV, equi-width histograms, and MCV frequencies. We also use native \textit{PostgreSQL} statistics collected from the original database as a reference. Because \texttt{ANALYZE} constructs statistics through sampling rather than exact counting, this reference captures the system's intrinsic estimation error and provides a practical baseline for evaluating synthesized data.

\begin{figure}[!t]
    \centering
    
    \includegraphics[height=0.55cm]{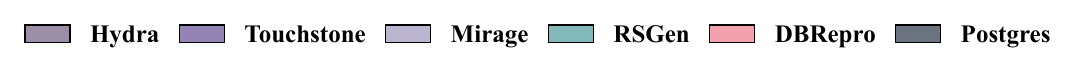}
    \vspace{-1mm} 

    \begin{subfigure}[b]{\linewidth}
        \centering
        \includegraphics[width=0.32\linewidth]{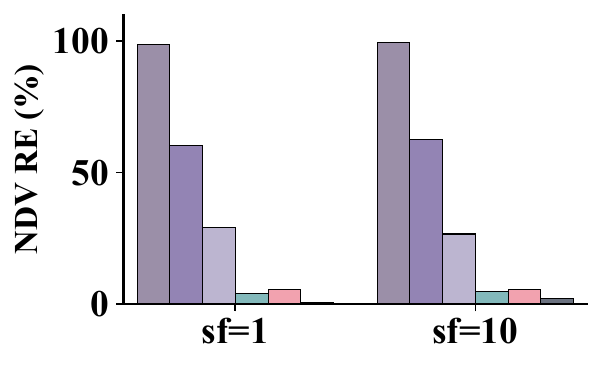}
        \hfill
        \includegraphics[width=0.32\linewidth]{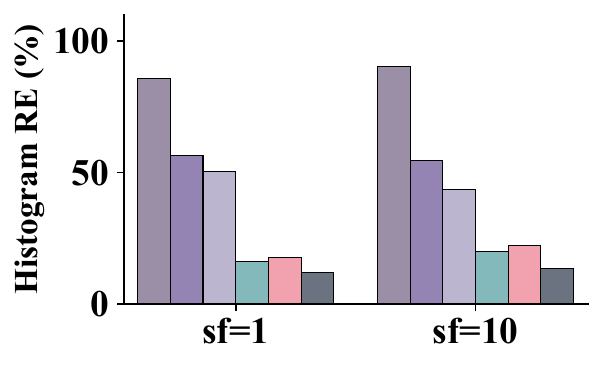}
        \hfill
        \includegraphics[width=0.32\linewidth]{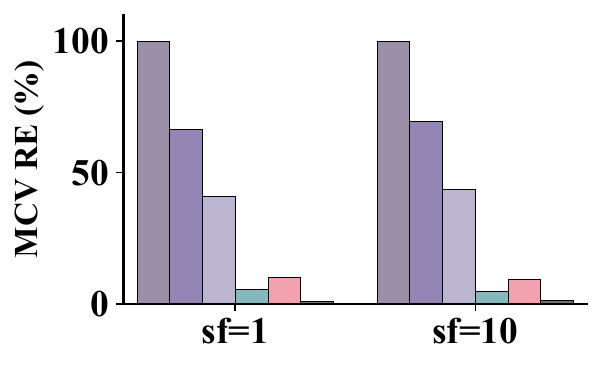}
        \vspace{-2mm}
        \caption{Relative errors of statistics for TPC-H (NDV, Histogram, MCV)}
        \label{fig:stat_err_tpch}
    \end{subfigure}
    
    \vspace{1mm} 
    
    \begin{subfigure}[b]{\linewidth}
        \centering
        \includegraphics[width=0.32\linewidth]{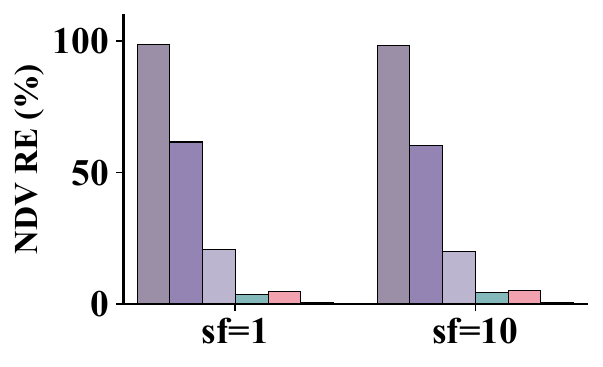}
        \hfill
        \includegraphics[width=0.32\linewidth]{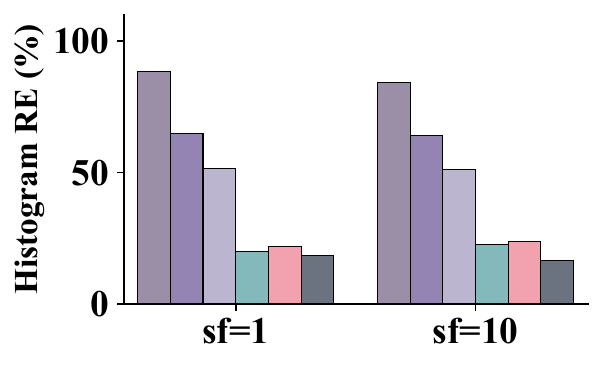}
        \hfill
        \includegraphics[width=0.32\linewidth]{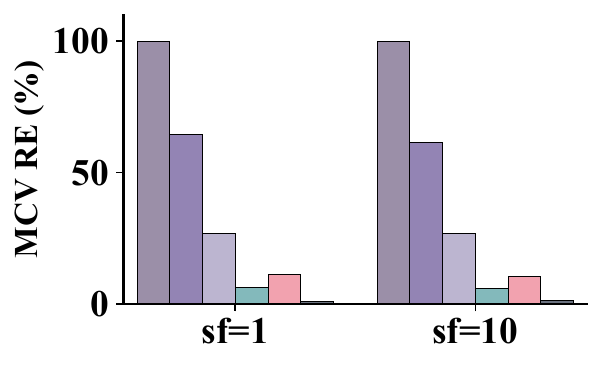}
        \vspace{-2mm}
        \caption{Relative errors of statistics for SSB (NDV, Histogram, MCV)}
        \label{fig:stat_err_ssb}
    \end{subfigure}
    
    \vspace{-3mm}
    \caption{Comparison of relative statistics errors between the synthetic and original databases.}
    \Description{Six grouped bar charts compare relative errors in the number of distinct values, equi-width histograms, and most-common-value frequencies. The top row shows TPC-H and the bottom row shows SSB, comparing workload-aware methods, RSGen, DBRepro, and the PostgreSQL reference.}
    \vspace{-4mm}
    \label{fig:statistics_recovery_overall}
\end{figure}

\begin{figure}[!t]
    \centering
    \includegraphics[height=0.55cm]{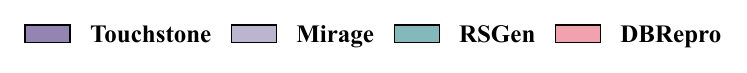}
    \vspace{-1mm} 
    
    \begin{subfigure}[b]{0.49\linewidth}
        \centering
        \includegraphics[width=\linewidth]{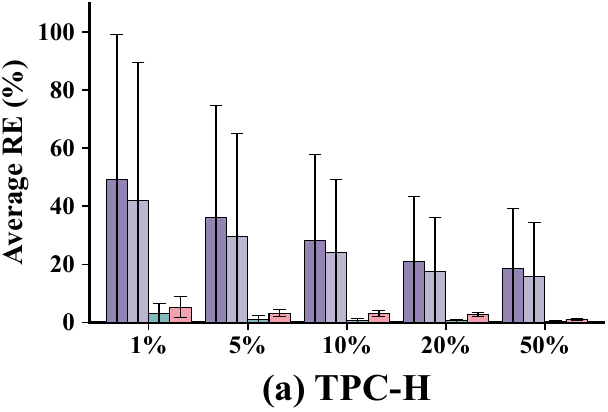}
        \label{fig:range_err_tpch}
    \end{subfigure}
    \hfill 
    \begin{subfigure}[b]{0.49\linewidth}
        \centering
        \includegraphics[width=\linewidth]{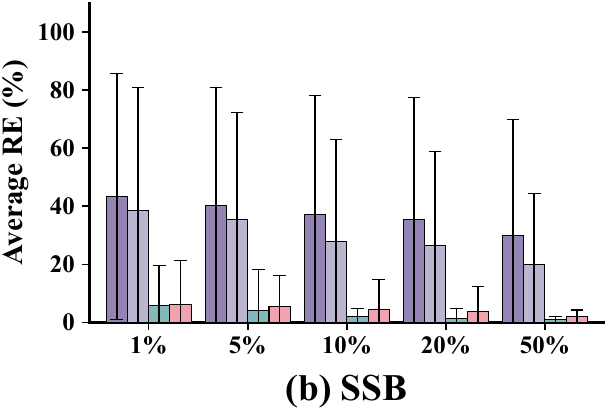}
        \label{fig:range_err_ssb}
    \end{subfigure}
    \vspace{-7mm}
    \caption{Average relative cardinality errors of range queries.}
    \Description{Two grouped bar charts show average range-query cardinality error on TPC-H and SSB at selectivities from one to fifty percent. DBRepro and RSGen remain substantially below the workload-aware methods across the selectivity levels.}
    \vspace{-4mm}
    \label{fig:range_query_error_overall}
\end{figure}

Figure~\ref{fig:statistics_recovery_overall} shows that workload-aware methods substantially distort the base distributions. Enforcing isolated query constraints without a global statistical anchor changes the distributions on which the optimizer's estimates depend. In contrast, \rsgen\ and \dbrepro\ approach the PostgreSQL sampling reference. They perform comparably on NDV and histograms, while \dbrepro's MCV error is 5\% higher on average. This deviation is an expected trade-off: to satisfy exact SCCs, \textit{Distribution Fusion} explicitly adjusts selected MCV frequencies when they conflict with the initial statistics. Unlike a purely data-driven reconstruction that cannot enforce these local cardinalities, this targeted adjustment preserves the distribution as closely as possible while satisfying the workload constraints.

Figure~\ref{fig:range_query_error_overall} further evaluates the reconstructed data density using range queries with selectivities from 1\% to 50\%. These queries directly probe histogram fidelity and the allocation of probability mass across value ranges. Consistent with the statistical metrics, \dbrepro\ and \rsgen\ outperform workload-aware methods at every selectivity level. The results confirm that \dbrepro\ retains sufficient global distribution fidelity to support optimizer decisions on unseen queries while enforcing the cardinalities required by the workload.

\subsection{RQ3: Practicality in Commercial DBMS}\label{rq3}

\begin{figure}[!thbp]
    \centering
    \includegraphics[width=0.98\linewidth]{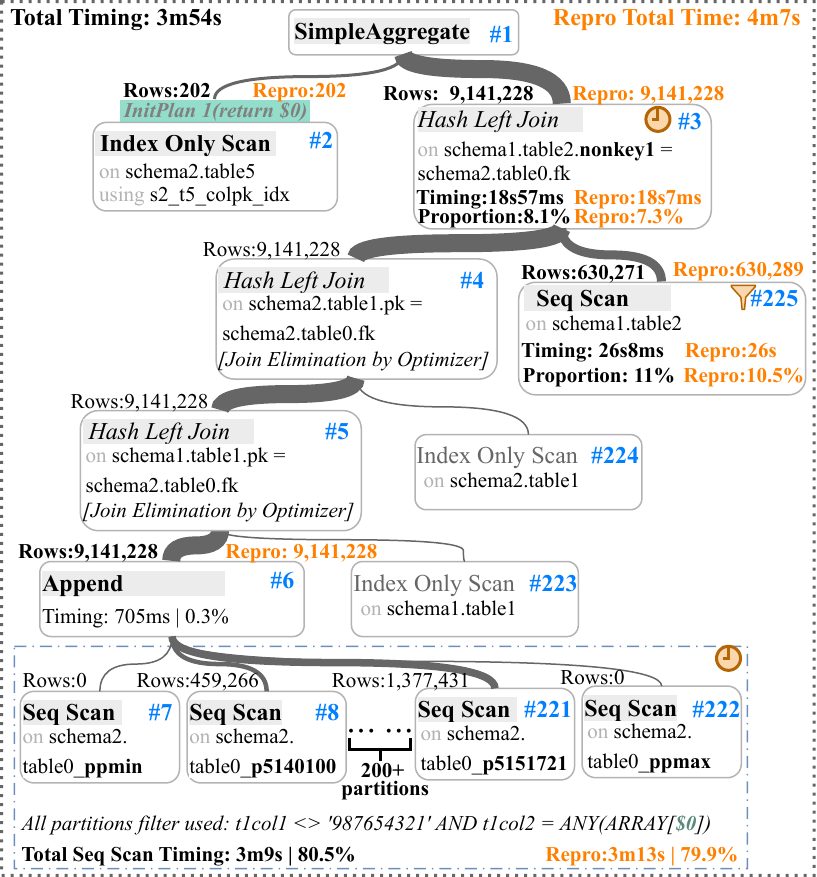} 
    \caption{Original physical execution plan of industrial slow query Q3, annotated with reproduction results.}
    \Description{A tree-shaped physical execution plan for industrial query Q3. Original row counts and operator times are shown with DBRepro results in orange, including scans, hash left joins, an append node, and a simple aggregate; the annotations show matching plan structure, cardinalities, and latency proportions.}
    \label{fig:industrial_qb}
\end{figure}

To answer RQ3, we evaluate \dbrepro's performance on commercial DBMSs in real-world industrial scenarios. Specifically, we deploy \dbrepro\ on KingbaseES to synthesize a nearly 1\,TB proxy database (serving over 20 million users) that contains 5 schemas, nearly 200 base tables, and over 5,000 physical partitions. The target workload includes six complex analytical slow queries extracted from the production environment, with an average execution time exceeding 100 seconds. We evaluate the reproduction fidelity using the same plan structural, cardinality, and latency metrics applied in RQ1. Table~\ref{tab:industrial_workload} reports the anonymized workload characteristics and per-query reproduction errors. Q3 is the case study shown in Figure~\ref{fig:industrial_qb}.

\begin{table}[!t]
    \centering
    \caption{Industrial workload and reproduction errors (\%). Tbl.: distinct base tables; Scan/Join: logical operators before optimizer elimination (partition scans grouped).}
    \label{tab:industrial_workload}
    \vspace{-2mm}
    \scriptsize
    \setlength{\tabcolsep}{1.5pt}
    \begin{tabular*}{\linewidth}{@{\extracolsep{\fill}}lrrrrrrrr@{}}
        \toprule
        & \multicolumn{5}{c}{\textbf{Workload characteristics}} & \multicolumn{3}{c}{\textbf{Reproduction error}} \\
        \cmidrule(lr){2-6}\cmidrule(l){7-9}
        \textbf{Q} & \textbf{DB} & \textbf{\#Tbl.} & \textbf{\#Scan} & \textbf{\#Join} & \textbf{Time} & \textbf{SCCs} & \textbf{JCCs} & \textbf{Latency} \\
        & \textbf{(GB)} & & & & \textbf{(s)} & $\text{RE}_{card}$ & $\text{RE}_{card}$ & $\text{RE}_{latency}$ \\
        \midrule
        Q1 & 80  & 1 & 2 & 0 & 175.6 & 0.00 & 0.00 & 7.07 \\
        Q2 & 140 & 6 & 6 & 5 & 45.7  & 0.00 & 3.43 & 0.72 \\
        Q3 & 207 & 5 & 5 & 3 & 234.0 & 0.00 & 0.00 & 1.91 \\
        Q4 & 91  & 6 & 8 & 3 & 10.8  & 0.64 & 0.00 & 3.34 \\
        Q5 & 47  & 6 & 6 & 5 & 84.5  & 0.00 & 0.00 & 1.43 \\
        Q6 & 142 & 4 & 4 & 3 & 62.1  & 0.00 & 16.20 & 13.80 \\
        \midrule
        \textbf{Avg.} & 117.8 & 4.67 & 5.17 & 3.17 & 102.1 & 0.11 & 3.27 & 4.71 \\
        \bottomrule
    \end{tabular*}
\end{table}

\subsubsection{End-to-End Reproduction Fidelity} Across all six real-world slow queries, which span 47--207\,GB of data, 1--6 base tables, 2--8 scans, 0--5 joins, and execution times of 10.8--234.0 seconds, \dbrepro\ reproduced identical plan structures. The SCC error remains below 0.64\% for every query, while the mean SCC, JCC, and latency-proportion errors are 0.11\%, 3.27\%, and 4.71\%, respectively. Q6 is the most challenging case because most of its joins are non-PK-FK. Precisely reproducing their fan-out would require exact key-frequency distributions for the pseudo-PK columns used by the Scaling Factor model, whose collection is costly on the production system. \dbrepro\ instead estimates scaling factors from lightweight statistics, trading some JCC accuracy for low-overhead extraction while retaining the identical plan structure. These results demonstrate consistent fidelity at both optimizer and operator levels across heterogeneous industrial workloads.

\subsubsection{Case Study: Execution Plan Analysis}

Figure~\ref{fig:industrial_qb} shows Q3's five-table plan: a four-table left-join tree plus a scalar \textit{InitPlan} lookup, running for nearly four minutes.
Reproducing this plan presents three major challenges. First, without consistent integrity constraints and global statistics, the optimizer may fail to trigger \textit{Join Elimination} (Nodes \#4 and \#5), causing structural divergence. Second, its SCCs span vastly different scales. Even a minor error in the tiny \textit{InitPlan} (\#2) can alter cost estimates, whereas the \textit{Seq Scan} on the 200-partition \textit{schema2.table0} must precisely filter 40 million rows to 9.1 million. This scan is the primary bottleneck, accounting for 80.5\% of total latency. Third, the final \textit{Hash Left Join} (\#3) imposes a non-key JCC (\textit{nonkey1}) requiring exactly 9.1 million output rows. Solvers designed only for PK-FK relationships cannot control the unpredictable fan-out of this join.

\dbrepro\ addresses these challenges by combining heuristic inference with \textit{Distribution Fusion}, achieving 0\% SCC error and exact input cardinalities for the table scans. Its Scaling Factor mechanism further eliminates fan-out error for the non-key JCC. Consequently, the synthetic latency distribution closely matches the original: all bottleneck operators contributing over 5\% of total latency remain below 1\% relative error. Preserving both the optimizer's structural decisions and the operator-level bottleneck profile demonstrates the practical value of the hybrid design for diagnosing industrial performance anomalies.

\subsection{Threats to Validity}

\noindent \textbf{Internal validity.} Threats concern the implementation of \dbrepro\ and the experimental measurements. Its analyzer produces constraint models for a comprehensive subset of relational operators, including unary, logical, and arithmetic selections as well as PK-FK and non-key joins. We used pair programming to check implementation correctness. Execution-latency measurements may also be affected by environmental noise; to reduce this error, we flushed the OS and DBMS caches before execution and reported the arithmetic mean of multiple runs.

\noindent \textbf{External validity.} Threats arise from the database systems selected for evaluation, since two systems cannot cover all RDBMSs. We chose PostgreSQL and KingbaseES to represent widely used open-source and commercial DBMSs driven by cost-based optimizers. Although this choice does not establish generality across every optimizer architecture, \dbrepro\ relies on standard catalog statistics used by CBOs rather than system-specific internals, supporting its applicability to other systems exposing comparable metadata.

%% file: 6-discussion.tex
\section{Related Work}

Database synthesis techniques fall into three paradigms:

\noindent\textbf{Independent Synthesis.} Independent synthesis methods construct databases from scratch without relying on existing production data. They are widely used to generate scalable datasets for standard performance benchmarks~\cite{tpchdbgen,tpcdsdsdgen}. For more flexible scenarios, statistical model-based tools let users specify data distributions and dependencies through configuration files~\cite{gray1994quickly,stephens2004mudd,ding2021dsb,DGL,PSDG,PDGF,myraid}, while commercial generators provide extensive user-defined generation capabilities~\cite{EMSDataGenerator,advanced_data_generator,mockaroo}. Recent systems further employ LLM agents to synthesize datasets from natural-language instructions~\cite{tonic_ai}. However, they cannot recover workload-specific characteristics from production.


\noindent\textbf{Data-driven Synthesis.} Data-driven approaches synthesize instances that preserve the characteristics of an existing source database. Metadata-based methods reconstruct data by inversely mapping catalog statistics~\cite{RSGen,shadowDB,10.1145/2723372.2735378,simple_realistic_data_gen,soltana2017synthetic}. Representative sample methods expand an authentic data sample to a target volume while maintaining its macroscopic distributions~\cite{buda2013cods,tay2013upsizer,buda2014vfds,ming2014bdgs,zhang2016dscaler,buda2017rex}. Additionally, probabilistic generative models with differential privacy synthesize relational data safely~\cite{patki2016synthetic,zhang2017privbayes,cai2023privlava,ge2024privacy}. This grounding yields more realistic global characteristics than independent generation. However, preserving global statistics alone does not guarantee exact intermediate cardinalities for a target workload.

\noindent\textbf{Workload-aware Synthesis.} Workload-aware methods synthesize databases explicitly guided by query workloads and execution requirements. Constraint-based methods formulate synthesis as a constraint satisfaction problem, generating tuples that satisfy specific intermediate-result sizes~\cite{binnig2007qagen,lo2014mybenchmark,arasu2011data,sanghi2018hydra,sanghi2018scalable,sanghi2022projection,3277355.3277411,li2024touchstone+,wang2024mirage,negi2023unshackling}. Coverage-based approaches automatically generate test databases and SQL inputs that satisfy query constraints and maximize objectives such as code or execution-path coverage~\cite{veanes2010qex,de2010constraint,suarez2010populating,suarez2017incremental,8453204,ren2020many}. Related DBMS testing techniques also synthesize SQL inputs for semantic differential testing or derive generation strategies from empirical bug patterns~\cite{10.14778/3712221.3712247,11222837}. Learning-based approaches use supervised autoregressive models to efficiently synthesize databases from large query workloads~\cite{yang2022sam}.

%% file: 7-conclusion.tex
\section{Conclusion}

We presented \dbrepro, an automated database synthesis framework that combines data-driven and workload-aware methodologies for offline slow query reproduction. By formulating database generation as a constrained distribution synthesis problem, \dbrepro\ progressively adjusts an initial global statistical distribution to satisfy exact local cardinality constraints. Its hybrid solver uses heuristic probabilistic inference and \textit{Distribution Fusion} for SCCs, together with a Scaling Factor-enhanced CP model for non-PK-FK joins. By reusing native DBMS slow-query records and non-intrusive catalog metadata, it avoids online query re-execution and base-table scans. This enables efficient runtime-context extraction whose cost is independent of the stored data volume, without accessing or exposing sensitive raw production tuples.

Extensive evaluations demonstrate high-fidelity reproduction across structural, cardinality, and latency metrics. Compared with data-driven baselines, \dbrepro\ reduces cardinality error by up to 20.3\%; compared with workload-aware methods, it lowers latency-proportion error by 21.5\% while reproducing 15\% more consistent execution plans. We further validated its practicality on a 1\,TB real-world industrial dataset managed by KingbaseES, demonstrating its utility for offline performance diagnosis.


%% file: 8-acknowledgments.tex
\begin{acks}
This work is supported by the National Natural Science Foundation of China under Grant Nos. 62472429, 62441230, the Fundamental Research Funds for the Central Universities, the Research Funds of Renmin University of China, Key Laboratory of Data Engineering and Knowledge Engineering MOE, and the A3 Foresight Program No. 62461146205. We gratefully acknowledge the support from the RUC Kingbase Database Collaborative Innovation Joint Laboratory.
\end{acks}

\clearpage

\section*{Data Availability Statement}

Our source code and benchmark datasets are publicly available~\cite{dbrepro_repo}. The real-world industrial dataset used in Section~\ref{rq3} cannot be publicly shared due to strict corporate data privacy regulations.